\documentclass{article}
\usepackage[english]{babel}
\usepackage{amsmath}

\catcode`\|=\active \def|{
\fontencoding{T1}\selectfont\symbol{124}\fontencoding{\encodingdefault}}
\newcommand{\nobracket}{}

\begin{document}

\title{Mandelstam-Leibbrandt prescription}

\author{J. Alfaro \\
Facultad de F\'\i sica, Pontificia Universidad Cat\'olica de Chile,\\
Casilla 306, Santiago 22, Chile.\\
jalfaro@uc.cl}

\maketitle

\begin{abstract}
  The light cone gauge is used frequently in string theory as well as gauge
  theories and gravitation. Loop integrals however have to be infrared 
  regulated to remove spurious poles. The most popular and consistent of these
  infrared regulators is the Mandelstam-Leibbrandt(ML) prescription. The
  calculations with ML are rather cumbersome, though. In this work we show
  that the ML can be replaced by a symmetry of the regulator. This symmetry
  simplify the calculations, reducing them to conventional dimensional
  regularization integrals.
\end{abstract}

\section{Introduction}

Computations in superstring theory as well as gauge theories, supersymmetry, 
gravitation and Chern-Simons theories  are often simplified by recurring to
the light cone gauge. The light cone gauge is termed one of the physical
gauges because ghosts decouple in these gauges{\footnote{There are some
subtleties related to  this point. Please see {\cite{Lbook}} chapter 4.4.}}.
To compute loop corrections in the light cone gauge has some
peculiarities,though: Spurious infrared poles appear, non local terms are
present  and Lorentz invariance is explicitly broken. To deal with these
problems an infrared regulator is needed. The most popular and internally
consistent regulator used at present is the Mandelstam-Leibbrandt regulator
(ML){\cite{man,lei}}. ML has very nice properties: The poles in the $k_{0}$
complex plane are situated such that the Wick's rotation from Euclidean to
Minkowsky space is justified; it preserves naive power counting of  loop
integrals; and in gauge theories, it maintains the Ward identities of the
gauge symmetry\cite{Lbook,soldati}.

Explicit computations with the ML are long and cumbersome, though.

Here we present a method to evaluate the loop integrals that appear in the
light cone gauge based on a scale symmetry of the regulator. No new integrals
are required, aside from the standard dimensionally regularized integrals. In
fact the ML prescription can be safely replaced by the scale symmetry and a
regularity condition. We do not have to specify the exact value of the two
null vectors of the ML, but merely its mutual relations. The results coincide
with the one obtained with ML.

\section{The new prescription}

Let us compute the following simple integral:
\[ A_{\mu} = \int d p  \frac{f ( p^{2} ) p_{\mu}  }{(n \cdot p)} \]
where $f$ is an arbitrary function.$d p$ is the integration measure in $d$
dimensional space and $n_{\mu}$ is a fixed null vector($(n \cdot n)=0$). This integral
is infrared divergent when $(n \cdot p)=0$.

The ML is:
\begin{equation}
     \frac{1}{(n\cdot p)} = \lim_{\varepsilon \rightarrow 0} \frac{(p\cdot \bar{n})}{(n\cdot p) (p\cdot
     \bar{n}) +i  \varepsilon} \label{ml}
   \end{equation} 
where $\bar{n}_{\mu}$ is a new null vector with the property $(n\cdot \bar{n}) =1$.

To compute $A_{\mu}$ we have to know the specific form of $f$, provide an
specific form of \ $n_{\mu}$ and $\bar{n}_{\mu}$, and evaluate the residues of
all poles of \ $\frac{f ( p^{2} )  }{(n\cdot p)}$ in the $p_{0}$ complex plane, a
rather formidable task for an arbitrary $f$.

Instead we want to point out the following symmetry:
\begin{equation}
  n_{\mu} \rightarrow \lambda  n_{\mu} , \bar{n}_{\mu} \rightarrow
  \lambda^{-1} \bar{n}_{\mu} , \lambda \neq 0, \lambda \varepsilon R
  \label{symmetry}
\end{equation}
It preserves the definitions of $n_{\mu}$ and $\bar{n}_{\mu}$:
\begin{eqnarray*}
  0=(n\cdot n) \rightarrow \lambda^{2} (n\cdot n)=0 &  & \\
  0= (\bar{n} \cdot \bar{n} )\rightarrow \lambda^{-2} (\bar{n} \cdot \bar{n} )=0 &  & \\
  1=(n\cdot \bar{n})\rightarrow (n\cdot \bar{n}) =1 &  & 
\end{eqnarray*}
We see from (\ref{ml}) that:
\[ \frac{1}{(n\cdot p)} \rightarrow \frac{1}{(n\cdot p)} \lambda^{-1} \]
Now we compute $A_{\mu}$, based on its symmetries. It is a Lorentz vector
which scales under (\ref{symmetry}) as $\lambda^{-1}$. The only Lorentz
vectors we have available in this case are $n_{\mu}$ and $\bar{n}_{\mu}$. But
(\ref{symmetry}) forbids $n_{\mu}$. That is:
\[ A_{\mu} =a  \bar{n}_{\mu} \]
Multiply by \ $n_{\mu}$ to find $(A\cdot n)=a$. Thus $a= \int d p f ( p^{2} )$.
Finally:
\[ \int d p  \frac{f ( p^{2} ) p_{\mu}  }{(n \cdot p)} = \bar{n}_{\mu} \int d p f (
   p^{2} ) \]
By the same token we find
\[ A_{\mu \nu \lambda} = \int d p  \frac{f ( p^{2} ) p_{\mu} p_{\nu}
   p_{\lambda}  }{(n \cdot p)} =a ( \bar{n}_{\mu} g_{\nu \lambda} )_{S} +b (
   \bar{n}_{\mu} \bar{n}_{\nu} n_{\lambda} )_{S} \]
where $( )_{S}$ means symmetric in all Lorentz indices.

We get:
\begin{eqnarray*}
  A_{\mu \nu \lambda} n^{\lambda} = \frac{1}{d} g_{\mu \nu} \int d p f ( p^{2}
  ) p^{2} 
\end{eqnarray*}

Therefore
\begin{equation*}
a= \frac{1}{d} \int d p f ( p^{2} ) p^{2}=-b
\end{equation*}

The integrals on $p_{\mu}$ are dimensionally regularized.

Therefore:
\[ \int d p  \frac{f ( p^{2} ) p_{\mu} p_{\nu} p_{\lambda}  }{(n \cdot p)} =
   \frac{1}{d} \int d p f ( p^{2} ) p^{2} \{ ( \bar{n}_{\mu} g_{\nu \lambda}
   )_{S} - ( \bar{n}_{\mu} \bar{n}_{\nu} n_{\lambda} )_{S} \} \]

\subsection{Generic integrals}

We consider now a more general integral. We will see here that regularity of
the answer will determine it uniquely.

Consider:
\begin{equation}
  A= \int d p  \frac{F ( p^{2} ,p\cdot q )}{(n \cdot p)} = (\bar{n} \cdot q )f ( q^{2} ,(n\cdot q )
  (\bar{n} \cdot q )) \label{generic}
\end{equation}
$q_{\mu}$ is an external momentum, a Lorentz vector. $F$ is an arbitrary
function. The last relation follows from (\ref{symmetry}), for a certain $f$
we will find in the following.

We get
\begin{eqnarray*}
  \frac{\partial A}{\partial q^{\mu}} = \int d p  \frac{F_{,u} p_{\mu}}{(n \cdot p)} =
  &  & \\
  \bar{n}_{\mu} f ( x,y ) +2 (\bar{n} \cdot q )q_{\mu} \frac{\partial}{\partial x} f
  ( x,y ) + [ ( \bar{n} \cdot q ) ^{2} n_{\mu} +(n\cdot q ) (\bar{n} \cdot q)  \bar{n}_{\mu} ]
  \frac{\partial}{\partial y} f ( x,y ) &  & 
\end{eqnarray*}

We defined $u=p\cdot q$,$x=q^{2}$,$y=(n\cdot q)  (\bar{n} \cdot q) $. $( )_{,u}$ means derivative
respects to $u$.
\begin{eqnarray}
  \frac{\partial A}{\partial q^{\mu}} n_{\mu} = \int d p F_{,u} = & g ( x ) =
  &  \nonumber\\
  f ( x,y ) +2 y \frac{\partial}{\partial x} f ( x,y ) +y
  \frac{\partial}{\partial y} f ( x,y ) &  &  \label{pdi1}
\end{eqnarray}
Assuming that the solution and its partial derivatives are finite in the
neighborhood of $y=0$, it follows from the equation that $f ( x,0 ) =g ( x )$.
That is the partial differential equation has a unique regular solution.

We will find the solution of (\ref{pdi1}) using the method of
characteristics{\cite{copson}}.
\begin{eqnarray*}
  \begin{array}{lll}
    \dot{x} =2y & \dot{y} =y & \\
    \dot{f} +f=g ( x ( t ) ) &  & \\
    y=C e^{t} & \dot{x} =2C e^{t} ,x=2C e^{t} +D & \\
    x-2y=D &  & 
  \end{array} &  & 
\end{eqnarray*}
The most general solution of the system is:
\begin{eqnarray*}
  f=b e^{-t} +e^{-t} \int_{- \infty}^{t} d t'  e^{t'} g ( x ( t' ) ) &  & 
\end{eqnarray*}
for $b$ arbitrary corresponding to the solution of (\ref{pdi1}) with
$g=0$(homogeneous solution). The regular solution of (\ref{pdi1}), $f_{0}$ ,
is obtained imposing that \ $b=0$. The reason being that the homogeneous
solution is: $f= \Pi ( x-2y )  y^{-1}$, with $\Pi$ an arbitrary function. We
readily see that $f$ will diverge at $y=0$, unless $\Pi ( x ) =0$, for all
$x$.

Moreover
\[ \lim_{t \rightarrow - \infty} f_{0} = \lim_{t \rightarrow - \infty} e^{-t}
   \int_{- \infty}^{t} d t'  e^{t'} g ( x ( t' ) ) =g ( D ) \]
That is $f_{0} ( x,0 ) =g ( x )$. $f_{0}$ is the unique regular solution of
(\ref{pdi1}).

What we have developed up to here shows that the scale transformation
(\ref{symmetry}) plus the regularity condition determines uniquely the value
of the integral (\ref{generic}).

\subsection{Application to loop integrals}

We consider now integrals that appear in gauge theory loops:

\begin{equation*}
  \int d p \frac{1}{[ p^{2} +2p\cdot q-m^{2} ]^{a}} \frac{1}{(n \cdot p)} = (\bar{n} \cdot q ) f
  ( x,y )  
\end{equation*}
In this case
\begin{equation*}
g ( x ) =-2a \int d p
  \frac{1}{[ p^{2} -x-m^{2} ]^{a+1}}
\end{equation*}
  Therefore
\begin{eqnarray*}
  f=e^{-t} \int_{- \infty}^{t} d t'  e^{t'} g ( x ( t' ) ) = &  & \\
  \int d p e^{-t} [ p^{2} -2 C e^{t'} -D-m^{2} ]^{-a} \frac{1}{-C} |_{-
  \infty}^{t} \nobracket = &  & \\
  - \frac{1}{y} \left\{ \int d p [ p^{2} -x-m^{2} ]^{-a} - \int d p [ p^{2}
  -x+2y-m^{2} ]^{-a} \right\} 
\end{eqnarray*}
We readily verify that  $f ( x,0 ) =-2a \int d p [ p^{2} -x-m^{2} ]^{-a-1} =g ( x )$

In the same way we get:
\begin{eqnarray*}
  \begin{array}{ll}
    \int d p \frac{1}{[ p^{2} +2p\cdot q-m^{2} ]^{a}} \frac{1}{( (n \cdot p) )^{2}} = (
    \bar{n} \cdot q )^{2} & f ( x,y )
  \end{array} &  & 
\end{eqnarray*}
with 
\[ f ( x,y ) = \frac{1}{y^{2}  } \int d p \{ [ p^{2} -x-m^{2} ]^{-a} - [ p^{2}
   -x+2y-m^{2} ]^{-a} -2a y [ p^{2} -x+2y-m^{2} ]^{-a-1} \} \]

Following the same procedure we  can get an answer for the whole family of
loop integrals:
\[ \int d p \frac{1}{[ p^{2} +2p\cdot q-m^{2} ]^{a}} \frac{1}{( (n \cdot p) )^{b}} = (
   \bar{n} \cdot q )^{b} ( -2 )^{b} \frac{\Gamma ( a+b )}{\Gamma ( a ) \Gamma ( b
   )} \int_{0}^{^{1}} d t t^{b-1} \int d p [ p^{2} -q^{2} +2n.q \overline{ n}
   .q t-m^{2} ]^{-a-b} \]
Using dimensional regularization, we obtain:
\begin{eqnarray}
  \int d p \frac{1}{[ p^{2} +2p\cdot q-m^{2} ]^{a}} \frac{1}{( (n \cdot p) )^{b}} = &  & 
  \nonumber\\
  ( -1 )^{a+b} i ( \pi )^{\omega} ( -2 )^{b} \frac{\Gamma ( a+b- \omega
  )}{\Gamma ( a ) \Gamma ( b )} ( \bar{n} \cdot q )^{b} \int_{0}^{^{1}} d t t^{b-1}
  \frac{1}{( m^{2} +q^{2} -2(n\cdot q )(\bar{ n} \cdot q) t )^{a+b- \omega}} , \omega
  =d/2 &  \label{I1}
\end{eqnarray}
We sketch the proof of equation (\ref{I1}).
\begin{equation*}
  \int d p \frac{1}{[ p^{2} +2p\cdot q-m^{2} ]^{a}} \frac{1}{( (n \cdot p) )^{b}} = (
  \bar{n} \cdot q )^{b} f ( b,a,x,y ) 
\end{equation*}
with
\begin{eqnarray}
  -2 a f ( b-1,a+1,x,y ) =b f ( b,a,x,y ) +2y \frac{\partial}{\partial x}  f (
b,a,x,y ) +y  \frac{\partial}{\partial y} f ( b,a,x,y ), &  &  \label{g1}\\
  f ( b,a,x,0 ) =- \frac{2a}{b} f ( b-1,a+1,x,0 ) &  &  \label{g2}
\end{eqnarray}

It is easy to check that (\ref{I1}) satisfies the partial differential
equation (\ref{g1}) and the boundary condition (\ref{g2}), so it is the unique
regular solution and thus determine the value of the integral.

Other integrals can be obtained deriving respects to $q_{\mu}$:
\begin{eqnarray}
  \int d p \frac{p_{\mu}}{[ p^{2} +2p\cdot q-m^{2} ]^{a+1}} \frac{1}{( (n \cdot p) )^{b}} =
  &  &  \nonumber\\
  \begin{array}{l}
    ( -1 )^{a+b} i ( \pi )^{\omega} ( -2 )^{b-1} \frac{\Gamma ( a+b- \omega
    )}{\Gamma ( a+1 ) \Gamma ( b )} ( \bar{n} \cdot q )^{b-1} b  \bar{n}_{\mu}
    \int_{0}^{^{1}} d t t^{b-1} \frac{1}{( m^{2} +x-2y t )^{a+b- \omega}} +\\
    ( -1 )^{a+b} i ( \pi )^{\omega} ( -2 )^{b} \frac{\Gamma ( a+b+1- \omega
    )}{\Gamma ( a+1 ) \Gamma ( b )} ( \bar{n} \cdot q )^{b} \int_{0}^{^{1}} d t
    t^{b-1} \frac{q_{\mu} -t ( n\cdot q \bar{n}_{\mu} + \bar{n} \cdot q n_{\mu} )}{(
    m^{2} +x-2y t )^{a+b+1- \omega}} 
  \end{array} &  & \label{I2}
\end{eqnarray}
and
\begin{eqnarray}
  \int d p \frac{p_{\mu}  p_{\nu}}{[ p^{2} +2p\cdot q-m^{2} ]^{a+2}} \frac{1}{( (n \cdot p)
  )^{b}} = ( -1 )^{a+b} i ( \pi )^{\omega} ( -2 )^{b-2} \{ \nobracket &  & 
  \nonumber\\
  \frac{\Gamma ( a+b- \omega )}{\Gamma ( a+2 ) \Gamma ( b-1 )} ( \bar{n} \cdot q
  )^{b-2} b  \bar{n}_{\mu} \bar{n}_{\nu} \int_{0}^{^{1}} d t t^{b-1}
  \frac{1}{( m^{2} +x-2y t )^{a+b- \omega}} &  &  \nonumber\\
  -2 \frac{\Gamma ( a+b+1- \omega )}{\Gamma ( a+2 ) \Gamma ( b )} ( \bar{n} \cdot q
  )^{b-1} b  \bar{n}_{\mu} \int_{0}^{^{1}} d t t^{b-1} \frac{( q_{\nu} -t (
  n\cdot q \bar{n}_{\nu} + \bar{n} \cdot q n_{\nu} ) )}{( m^{2} +x-2y t )^{a+b+1-
  \omega}} &  &  \nonumber\\
  -2 \frac{\Gamma ( a+b+1- \omega )}{\Gamma ( a+2 ) \Gamma ( b )} ( \bar{n} \cdot q
  )^{b-1} b \bar{n}_{\nu} \int_{0}^{^{1}} d t t^{b-1} \frac{q_{\mu} -t ( n\cdot q
  \bar{n}_{\mu} + \bar{n} \cdot q n_{\mu} )}{( m^{2} +x-2y t )^{a+b+1- \omega}} & 
  & + \nonumber\\
  4 \frac{\Gamma ( a+b+2- \omega )}{\Gamma ( a+2 ) \Gamma ( b )} ( \bar{n} \cdot q
  )^{b} \int_{0}^{^{1}} d t t^{b-1} \frac{[ q_{\mu} -t ( n\cdot q \bar{n}_{\mu} +
  \bar{n} \cdot q n_{\mu} ) ] [ q_{\nu} -t ( n\cdot q \bar{n}_{\nu} + \bar{n} \cdot q n_{\nu}
  ) ]}{( m^{2} +x-2y t )^{a+b+2- \omega}} & \} \nobracket &  \label{I3}
\end{eqnarray}
The right hand side of equations (\ref{I1},\ref{I2},\ref{I3}) is analytic in
the parameters $a,b, \omega$ almost everywhere in their respective complex
planes, so it provides the analytic extension of the integral to these wider
domains.

\section{Comparison with ML}

The simpler integral $A_{\mu} ,A_{\mu \nu \lambda}$ of section 2, agree with
the ML prescriptions{\cite{Lbook}}.

But, in this section we want to compute a more involved integral, in order to
compare both finite and divergent results with ML's.

We want to compute:
\[ A ( \sigma ,q ) = \int d p  \frac{( p^{2} )^{\sigma -1}}{( p-q )^{2} ( (n \cdot p)
   )^{2}} \]
We introduce Feynman parameters {\cite{peskin}}
\[ \frac{1}{A_{1}^{m_{1}} A_{2}^{m_{2}}} = \int_{0}^{1} d x  \frac{x^{m_{1}
   -1} ( 1-x )^{m_{2} -1}}{[ x A_{1} + ( 1-x ) A_{2} ]^{m_{1} +m_{2}}}
   \frac{\Gamma ( m_{1} +m_{2} )}{\Gamma ( m_{1} ) \Gamma ( m_{2} )} \]
to get,
\[ A ( \sigma ,q ) = ( 1- \sigma ) \int_{0}^{1} d x \int d p \frac{x^{-
   \sigma}}{[ p^{2} + ( 1-x ) ( -2p\cdot q+ \bar{q}^{2} ) ]^{2- \sigma}( (n \cdot p)
   )^{2}} \]
Using (\ref{I1}), we finally get
\begin{eqnarray*}
  A ( \sigma ,q ) =4 ( \bar{n} \cdot q )^{2} ( -1 )^{\sigma} i ( \pi )^{\omega}
  \frac{\Gamma ( 4- \sigma - \omega )}{\Gamma ( 1- \sigma )} \int_{0}^{1} d x
  x^{- \sigma} ( 1-x )^{\sigma + \omega -2} &  & \\
  \int_{0}^{^{1}} d t t [ q^{2} x+2 t (n\cdot q)( \bar{ n} \cdot q )( 1-x ) ]^{\sigma +
  \omega -4} &  & 
\end{eqnarray*}

This coincides with {\cite{leib1986}} equation C4, by the change of variable
$t=1-y$ and $\sigma = \omega$.

Notice that we have considered $\bar{q}^{2} \neq q^{2}$ and taken the limit
$\bar{q}^{2} =q^{2}$ after evaluating the integral. This is justified because
the integral is a regular function of $q^{2}$. If we expand $A ( \sigma ,q )$
in powers of $q^{2}$, each term of the series can be evaluated using
(\ref{I1}). The summation of the series is equivalent to the procedure we
followed above.

\section{Conclusions}

We have developed a way of evaluation of the light cone loop integrals based
on the scale symmetry (\ref{symmetry}) and the condition of regularity of the
solution.We do not have to specify the exact value of the two null vectors of the ML, but merely its mutual relations.  The answer is the same than in
the ML prescription, but a significant simplification of the calculation is
available now.

For future work, we want to  mention that the scale transformation
(\ref{symmetry}) is also a symmetry of the uniform prescription introduced by
Leibbrandt{\cite{Lbook}} to treat the spurious infrared poles in light-cone,
axial, planar and temporal gauge. The application of the method  presented
here to these more general gauges will be done elsewhere.

\section{Acknowledgements}

The work of J. A. is partially supported by Fondecyt 1150390, CONICYT-PIA-ACT1102
and CONICYT-PIA-ACT1417.

\end{document}